\documentclass[12pt,preprint]{aastex}

\shorttitle{\indent \def Nascent solar wind observed by EIS}
\shortauthors{Tian et al.}

\begin{document}

\title{The nascent fast solar wind observed by EIS onboard Hinode}

\author{Hui Tian\altaffilmark{1}, Chuanyi Tu\altaffilmark{1}, Eckart Marsch\altaffilmark{2}, Jiansen He\altaffilmark{2},
Suguru Kamio\altaffilmark{2}}

\altaffiltext{1}{School of Earth and Space Sciences, Peking
University, 100871 Beijing, China; tianhui924@gmail.com}

\altaffiltext{2}{Max-Planck-Institut f\"ur Sonnensystemforschung,
37191 Katlenburg-Lindau, Germany}

\begin{abstract}
The origin of the solar wind is one of the most important unresolved
problems in space and solar physics. We report here the first
spectroscopic signatures of the nascent fast solar wind on the basis
of observations made by the EUV Imaging Spectrometer (EIS) on Hinode
in a polar coronal hole, in which patches of blueshift are clearly
present on dopplergrams of coronal emission lines with a formation
temperature of $\lg(T/\rm{K})\geq$6.0. The corresponding upflow is
associated with open field lines in the coronal hole, and seems to
start in the solar transition region and becomes more prominent with
increasing temperature. This temperature-dependent plasma outflow is
interpreted as evidence of the nascent fast solar wind in the polar
coronal hole. The patches with significant upflows are still
isolated in the upper transition region but merge in the corona, in
agreement with the scenario of solar wind outflow being guided by
expanding magnetic funnels.
\end{abstract}

\keywords{Sun: corona---Sun: transition region---Sun: solar
wind---Sun: UV radiation}

\section{Introduction}

As one of the most important unresolved problems in space and solar
physics, the origin of the solar wind has received much attention
and was studied intensively since the discovery of the solar wind.
While it is well established that coronal holes are the source
regions of the fast solar wind
\citep[e.g.,][]{Krieger1973,Bell1976,McComas1998,Kohl2006}, the
sources of the slow solar wind remained unclear. Several possible
solar locations have been suggested, including boundaries of polar
coronal holes and helmet streamers
\citep[e.g.,][]{Gosling1981,Wang1990,Antonucci2006,Noci2007}, small
open-field regions embedded in the quiet Sun
\citep{He2007,Tian2008a,He2009,Tian2009}, and edges of active
regions \citep{Kojima1999,Sakao2007,Marsch2008}.

Possible spectroscopic signatures of the nascent fast solar wind
have, in particular on the basis of observations made by the Solar
Ultraviolet Measurements of Emitted Radiation
\citep[SUMER,][]{Wilhelm1995,Lemaire1997} instrument in SOHO, been
intensively studied and identified in the past decade. Especially,
the Ne~{\sc{viii}}~($\lambda770$) line, formed in the upper
transition region (TR) at approximately $\lg(T/\rm{K})=$5.8, was
found to be blueshifted on average in coronal holes
\citep{Dammasch1999,PeterJudge1999,Wilhelm2000,Xia2003,Aiouaz2005,McIntosh2007,Tian2008b,Tian2008c}.
Two-dimensional images of the Doppler shift of Ne~{\sc{viii}}
revealed blue shifts of 3-6~km/s along the network lanes in polar
coronal holes \citep{Hassler1999,Wilhelm2000}. These patches of blue
shift were found to be associated with coronal magnetic funnels, and
were interpreted as the initial outflow of the fast solar wind
\citep{Tu2005}.

Due to the lack of strong coronal emission lines in the SUMER spectra,
two-dimensional dopplergrams of emission lines with a formation temperature
higher than that of Ne~{\sc{viii}} could not be derived in coronal holes.
Yet recent observations with the EUV Imaging Spectrometer \citep[EIS,][]{Culhane2007}
onboard Hinode revealed clear blue shifts at the edges of active regions.
The speed of this upflow increases with increasing temperature, from about
3~km/s at $\lg(T/\rm{K})=$5.8 to 28~km/s at $\lg(T/\rm{K})=$6.3 \citep{DelZanna2008}.
This steady plasma outflow was interpreted as an indicator of the nascent
slow solar wind \citep{Harra2008}. A similar study of the velocity structures
in coronal holes, however, is hampered significantly by the weakness of the
emission there, and thus spectroscopic signatures of the nascent fast solar
wind in coronal holes have not yet been identified by using EIS data.

In this paper, we will present such results derived from a data set obtained
by EIS with a very long exposure time of 150~s, and thus we can report here
the first identification of spectroscopic signatures of the nascent fast
solar wind from EIS observations in a polar coronal hole.

\section{Data reduction and analysis}

The EIS data analysed here were acquired at the northern polar
region from 14:13 to 18:17 UT on October 10, 2007. This data set was
previously analysed by \cite{Banerjee2009} to search for signatures
of Alfv\'{e}n waves. The combination of using the $2^{\prime\prime}$
slit and a very long exposure time of 150~s greatly improved the
count statistics, and thus allows for a reliable derivation of the
Doppler shifts from the line profiles. The raster increment was
about $2^{\prime\prime}$, and 101 exposures were made during the
scan, so that the scanned region has a size of about
$200^{\prime\prime}\times512^{\prime\prime}$. The wavelengths and
formation temperatures of the selected lines are listed in
Table~\ref{table1}.

\begin{table}[]
\caption[]{Emission lines used for this study. Here $\lambda$ represents the
wavelength and $T$ the formation temperature of the emission line.}
\label{table1}
\begin{center}
\begin{tabular}{p{1.0cm} p{1.2cm} p{1.5cm}| p{1.0cm} p{1.1cm} p{1.5cm}}
\hline Ion & $\lambda$ ({\AA}) &  $\lg(T/\rm{K})$
     & Ion & $\lambda$ ({\AA}) &  $\lg(T/\rm{K})$ \\
\hline
He~{\sc{ii}}  & 256.32 &   4.70 &Fe~{\sc{x}}   & 184.54  &  6.00 \\
O~{\sc{v}}    & 192.90 &   5.40 &Fe~{\sc{xii}} & 195.12  &  6.11 \\
Mg~{\sc{vi}}  & 270.39 &   5.60 &Fe~{\sc{xiii}}& 202.04  &  6.20 \\
Fe~{\sc{viii}}& 185.21 &   5.60 &Fe~{\sc{xiv}} & 270.52  &  6.25 \\
Si~{\sc{vii}} & 275.35 &   5.80 &Fe~{\sc{xv}}  & 284.16  &  6.30 \\
\hline
\end{tabular}
\end{center}
\end{table}

The standard EIS correction routine \textit{eis\_prep.pro} available
in the SolarSoft (SSW) package was adopted to calibrate the data,
including subtraction of the dark current, removal of cosmic rays
and hot pixels, and radiometric calibration. The amount of slit tilt
has been estimated by averaging the velocities along each row from a
large raster of the quiet Sun taken early in the
mission~\citep{Mariska2007}. It was removed by applying the SSW
routine \textit{eis\_slit\_tilt.pro}. To correct for the orbital
variation caused by thermal effects on the instrument, we first
summed up all profiles of the strong line Fe~{\sc{xii}}~195.12 in
each exposure, and then calculated the spectral position of the line
center for each summed profile. The resulting temporal evolution of
the line center position was used to eliminate the orbital variation
in all lines. The corrections of the slit tilt and orbital variation
were applied to the wavelength vector at each spatial location of an
image before further data processing.

By calculating the cross correlation between the intensity images of the
lines He~{\sc{ii}}~256.32 and O~{\sc{v}}~192.90, we found an offset by
$17^{\prime\prime}$ in the north-south direction between the images in these
two EIS wavelength bands. We therefore chose the common regions where the
observations in different lines overlap, and subsequently reduced the size of
the studied region to $200^{\prime\prime}\times478^{\prime\prime}$. Then we
averaged the resulting data over two pixels along the slit to improve the
signal-to-noise ratio and made the pixel size comparable in both dimensions
of solar-X and solar-Y. Finally, a running average over 3 pixels in both
spatial dimensions was applied to the data, which again greatly improved the
signal-to-noise ratio.

\begin{figure*}
\centering
\includegraphics{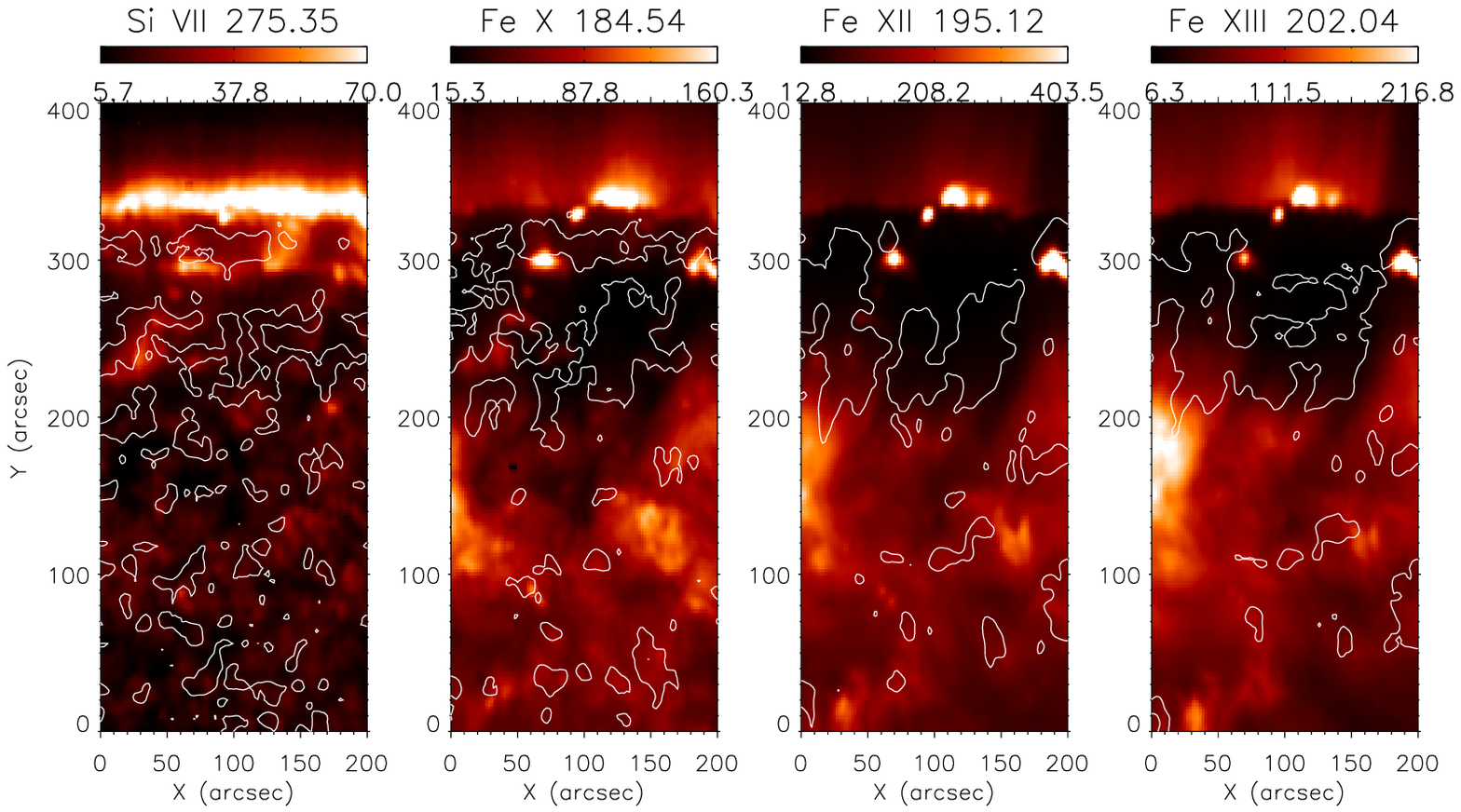}
\includegraphics{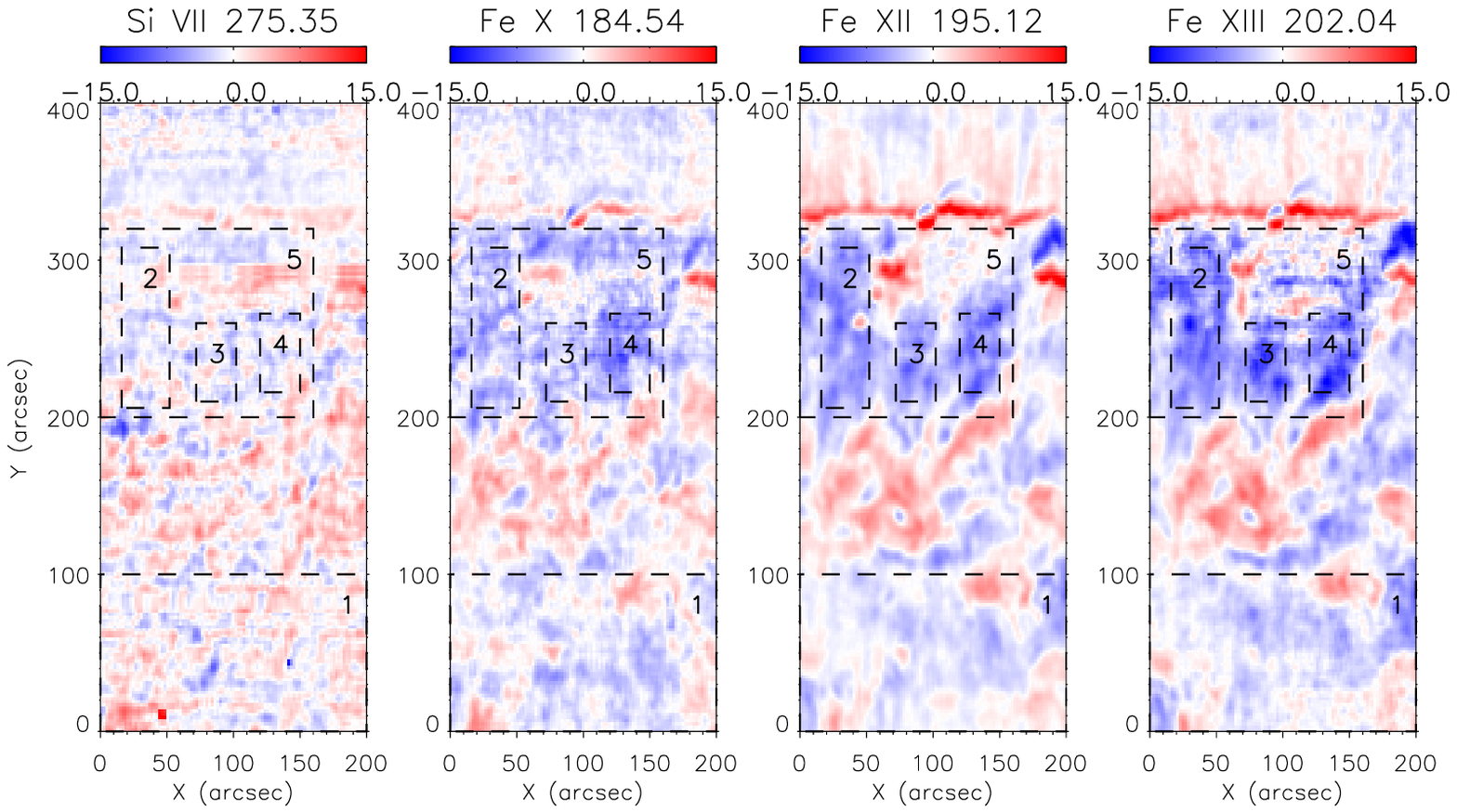}
\caption{~Maps of radiance (upper panels) in arbitrary units and Doppler
shift (lower panels) in colour-coded units of km/s for four emission lines.
Contours enclosing the lowest 20\% percent of the Doppler values are over
plotted on the radiance map of the corresponding line. The dashed lines on
the dopplergrams outline the rectangular regions in which the profiles of
each line as listed in Table~\ref{table1} were accumulated and further
studied.} \label{fig.1}
\end{figure*}

A single gaussian fit was then applied to each line profile that is
clean from blends. According to \cite{Young2007}, the
Fe~{\sc{viii}}~185.21, Si~{\sc{vii}}~275.35, Fe~{\sc{x}}~184.54,
Fe~{\sc{xii}}~195.12, and Fe~{\sc{xiii}}~202.04 line profiles in our
data can all be approximated well by a single gaussian. The
Fe~{\sc{viii}}~185.21 line is blended with the Ni~{\sc{xvi}}~185.23
line in active regions, but in the quiet Sun and coronal holes this
blend is almost absent. The Mg~{\sc{vi}}~270.39 and
Fe~{\sc{xiv}}~270.52 lines are blended with each other, and
therefore we used a double-gaussian fitting procedure to separate
them. The Fe~{\sc{xv}}~284.16 line is weakly blended with
Al~{\sc{ix}}~284.03, and a double-gaussian fit was applied for their
decomposition. There are several lines (namely O~{\sc{v}}~192.80,
Fe~{\sc{xi}}~192.83, Ca~{\sc{xvii}}~192.82) on the blue wing of
O~{\sc{v}}~192.90, and thus we also had to use a double-gaussian fit
to extract the line profile of O~{\sc{v}}~192.90, thereby treating
the three blends as one gaussian component, respectively. The
He~{\sc{ii}}~256.32 line and its blends are difficult to separate,
and thus we did not use this line for our spectroscopic study. The
line radiance and central position of every profile were thus
obtained from the fits. However, although we have improved the
signal-to-noise ratio a lot, there are still many profiles which can
not be well fitted, especially those of weak lines. Thus, in
Figure~\ref{fig.1} we only present maps of radiance and Doppler
shift for four strong lines and exclude the northern most part of
the field of view (FOV).

\begin{figure}
\centering {\includegraphics[width=\textwidth]{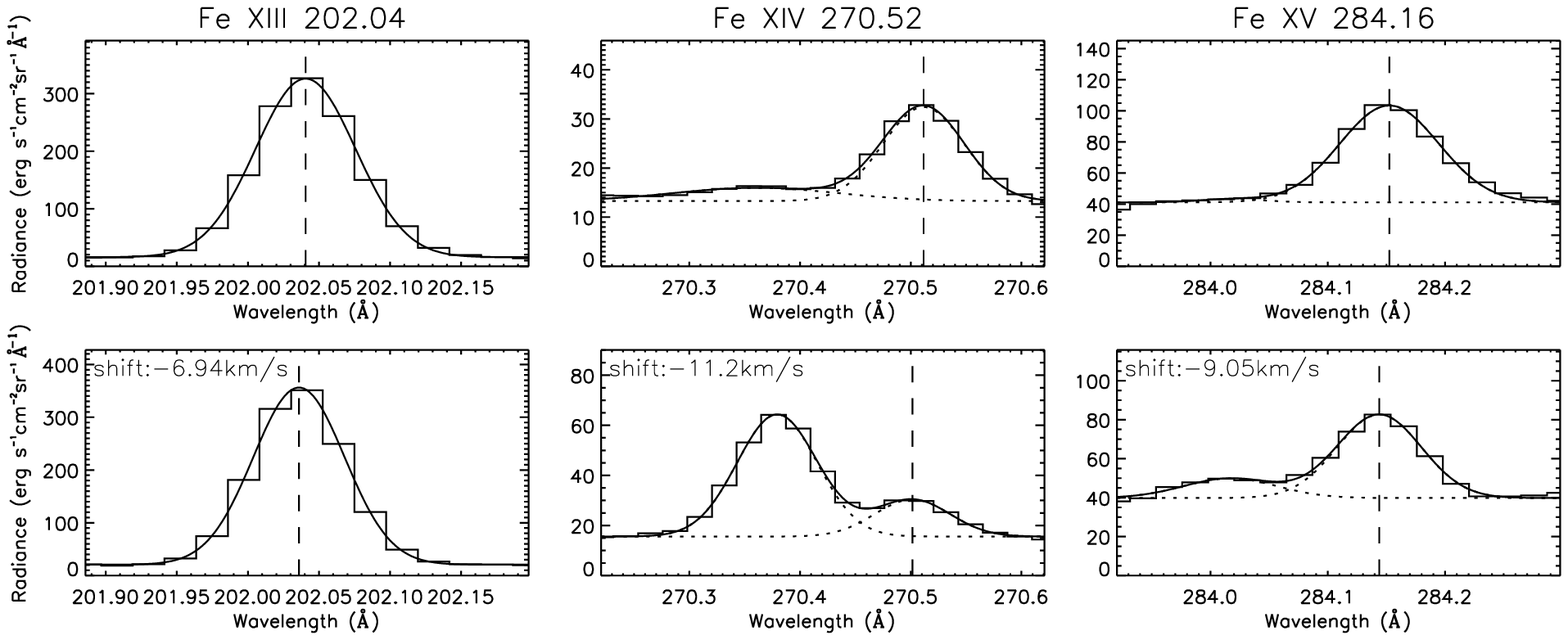}}
\caption{Averaged line profiles at positions between $65^{\prime\prime}$
and $143^{\prime\prime}$ above the limb (upper) and region 3 (lower) for
three emission lines. The observed and fitted line profiles are represented
by the histograms and solid curves, respectively. For the blended lines, the
two gaussian components are shown as two dotted curves. The vertical dashed
lines indicate the central spectral positions. The absolute Doppler shift
as revealed from the averaged profile in region 3 is also shown for each line.}
\label{fig.2}
\end{figure}

We further selected several subregions from the FOV, which are
outlined by dashed rectangles in Figure~\ref{fig.1}. Regions 1 and 5
correspond to the quiet Sun and coronal hole, respectively. Regions
2-4 are small sub-regions where significant blue shift of
Fe~{\sc{xiii}} is present inside the coronal hole. For each line,
the profiles in each region were averaged. These profiles are all
good enough to permit reliable gaussian fits. Several examples of
the averaged line profiles are presented in Figure~\ref{fig.2}.

\begin{figure}
\centering {\includegraphics[width=\textwidth]{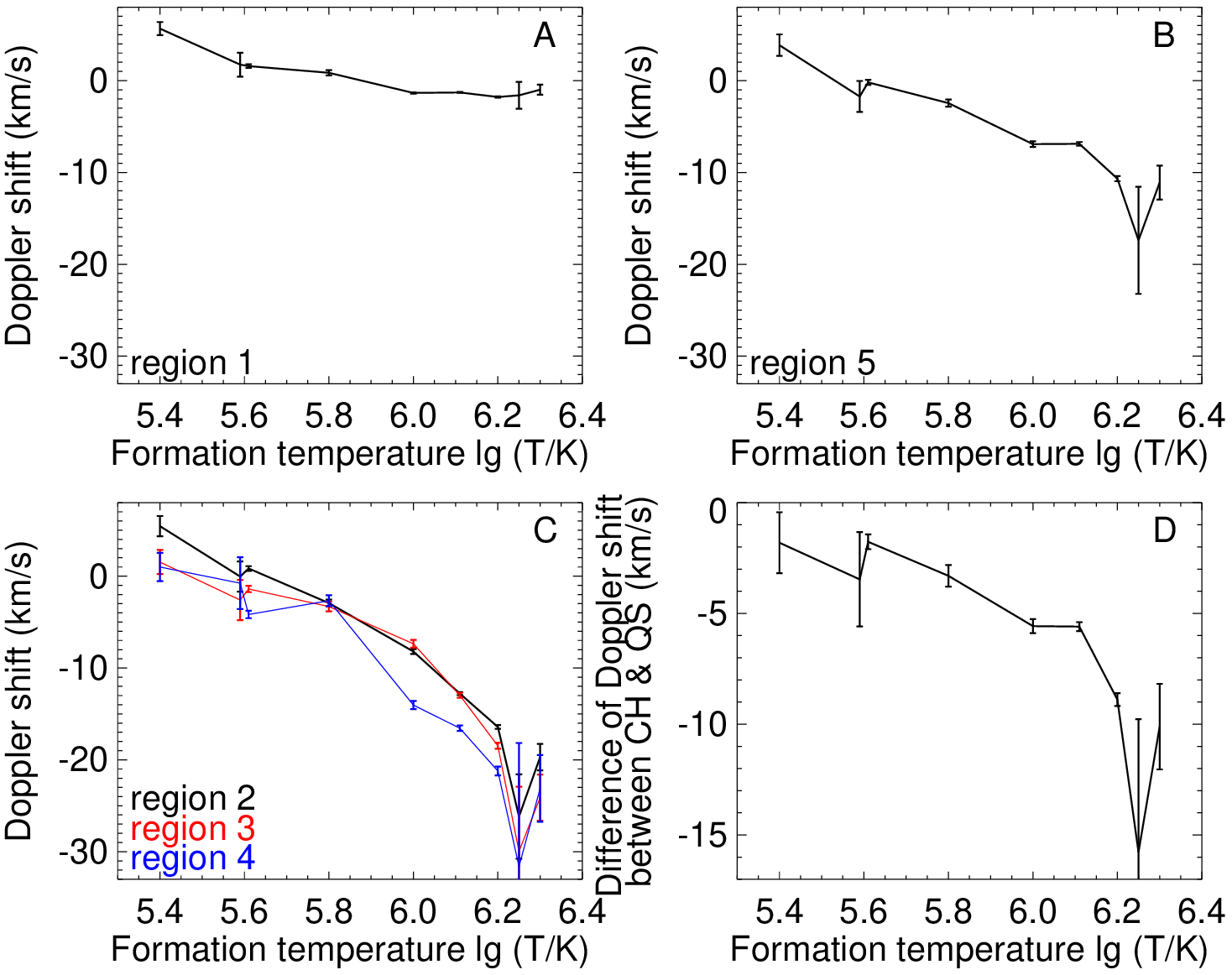}}
\caption{A-C: Temperature dependence of the Doppler shift for various
profiles averaged over several selected regions. The error bars
represent the corresponding fitting errors of the line centers. D:
Temperature dependence of the difference between the Doppler shifts for
the averaged profiles in the coronal hole (region 5) and quiet Sun
(region 1). The error bars represent the uncertainties originating from
the corresponding fitting errors of the line centers. To avoid overlapping,
the data points for the Mg~{\sc{vi}} and Fe~{\sc{viii}} lines are plotted
at $\lg(T/\rm{K})=$5.59 and $\lg(T/\rm{K})=$5.61, respectively. }
\label{fig.3}
\end{figure}

The rest wavelengths were determined by assuming a zero shift of the
accumulated emission at positions between $65^{\prime\prime}$ and
$143^{\prime\prime}$ above the limb (the northern most part of the
FOV). Here the limb refers to the radial position where the radiance
of Si~{\sc{vii}}~275.35 maximizes. This assumption is reasonable
because non-radial velocities usually cancel in the corona
\citep{Kamio2007}, and dynamic events are very rare in such high
altitudes. In this way we determined the rest wavelength of each
line, and thus obtained the absolute Doppler shift. The emission of
the Mg~{\sc{vi}}~270.39 and O~{\sc{v}}~192.90 lines was very weak
between $65^{\prime\prime}$ and $143^{\prime\prime}$ above the limb,
and consequently the fitting errors were very large. From Chianti
\citep{Dere1997,Landi2006} we know the spectral distance between
Mg~{\sc{vi}}~270.39 and Fe~{\sc{xiv}}~270.52. So the rest wavelength
of Mg~{\sc{vi}}~270.39 could be calculated by subtracting this
spectral distance from the rest wavelength of the
Fe~{\sc{xiv}}~270.52 line. But we could not use this method for the
O~{\sc{v}}~192.90 line, since its blends were too complicated.
Instead, we simply assumed a net red shift of 6~km/s in the quiet
Sun (region 1), following the statistical result in Figure~8 of
\cite{Xia2004}.

The absolute Doppler shift of each line in different selected
regions are presented in Figure~\ref{fig.3}. Here positive and
negative values correspond to red and blue shift, respectively. In
Figure~\ref{fig.3}D we present the temperature dependence of the
difference between Doppler shift for the averaged profiles in the
coronal hole (region 5) and quiet Sun (region 1). The Doppler shifts
in regions 1-5 have been divided by 0.71, 0.37, 0.44, 0.43, and
0.37, respectively, to account for the line-of-sight effect under
the assumption of radial outflows.

\section{Results and Discussion}

In the past 10 years, SUMER observations have contributed a lot to our
understanding of the origin of the fast solar wind. Dopplergrams of the
upper-TR line Ne~{\sc{viii}}~($\lambda770$) revealed prevailing blue shifts
in coronal holes, which was interpreted as a signature of the nascent fast
solar wind. However, because of the lack of strong hotter lines in the SUMER
spectra, higher-temperature (e.g., $\lg(T/\rm{K})\geq$6.0) spectroscopic
information on the initial outflow of the fast solar wind was still missing.

\begin{figure}
\centering {\includegraphics[width=0.5\textwidth]{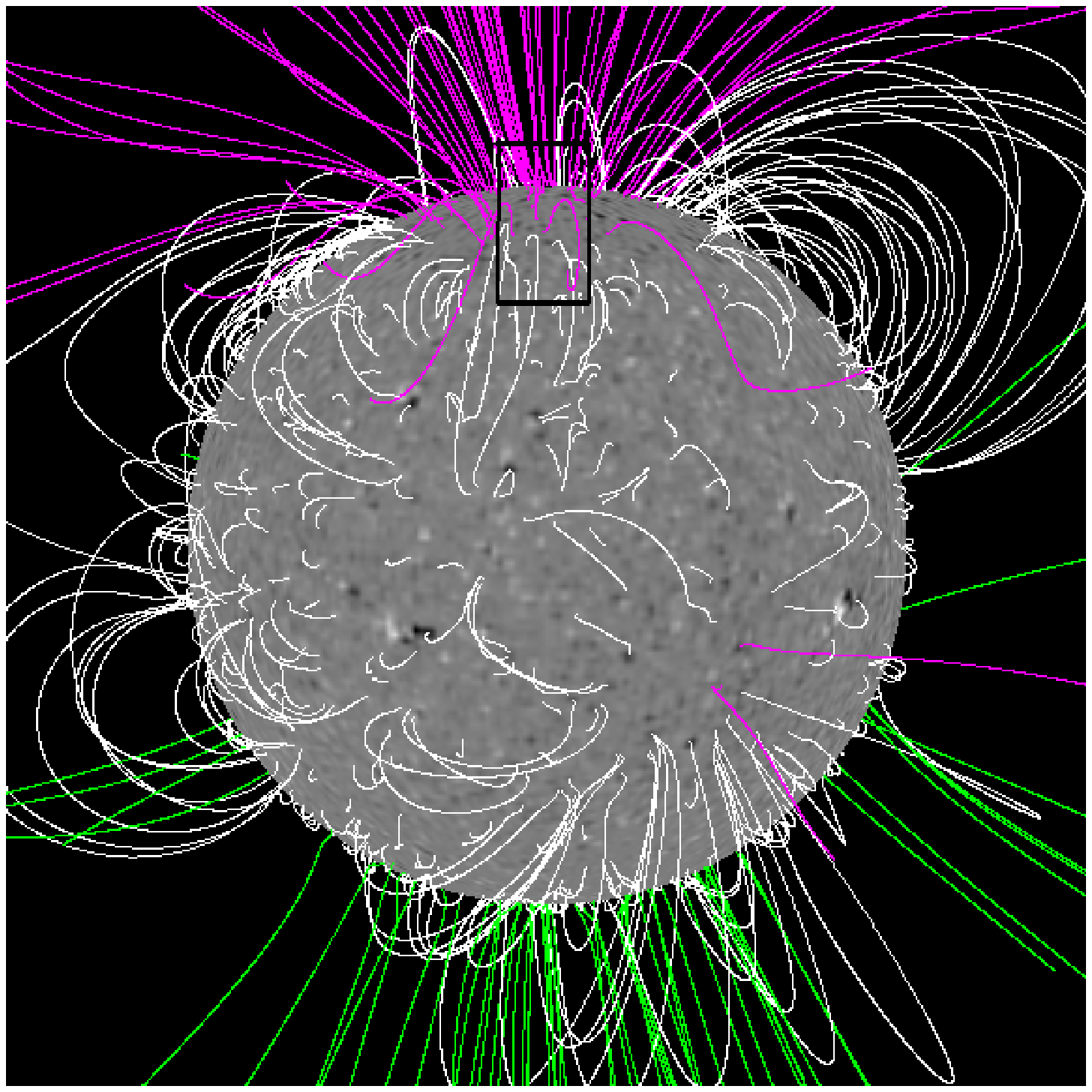}}
\caption{Magnetic field structure of the Sun at 18:06 on October 10,
2007, obtained by using the PFSS package available in SSW. Open
field lines with different polarities passing through the source
surface at 2.5 solar radii are illustrated in red and green,
respectively. Closed field lines are shown in white. The rectangle
outlines the observation region of EIS. } \label{fig.4}
\end{figure}

By analysing the spectra acquired by EIS, we have managed to produce
new two-dimensional images of the Doppler shift in a coronal hole
for several emission lines with formation temperatures higher than
that of Ne~{\sc{viii}}. Patches of significant blue shift are
clearly present in the coronal hole on the dopplergrams of coronal
emission lines with a formation temperature higher than
$\lg(T/\rm{K})=$6.0. At some locations the blue shift can reach
$\sim$20~km/s. While outside the coronal hole there is mainly red
shift and weak blue shift, as can be seen in Figure~\ref{fig.1}. In
Figure~\ref{fig.4} we illustrate magnetic field lines which were
obtained by using the PFSS (potential field source surface)
model~\citep{Schrijver2001}. It is very clear that the significant
blue shifts in the coronal hole are associated with open field
lines, and thus should be an indicator of the solar wind outflow.

Several coronal bright points (BPs) are also present in the coronal hole
studies here. However, all of these bright points reveal blue shifts on one
side and red shifts on the other side, which we interpret as spectroscopic
signatures of siphon flows along the loops associated with the BPs. These
adjacent downflows and upflows might also be signatures of the outward flows
that are produced by magnetic reconnection in BPs
\citep{Brosius2007,Tian2008d}.

The outflow in the coronal hole seems to start in the TR. The dopplergram of
the Si~{\sc{vii}} line reveals some patches of weak blue shift in the coronal
hole. There are more blue-shift patches inside the coronal hole than outside.
We notice that the formation temperature of Si~{\sc{vii}} is similar to
Ne~{\sc{viii}}. However, the blue shift of Si~{\sc{vii}} in this coronal hole
seems to be less prominent than that of Ne~{\sc{viii}} in the polar coronal
holes observed by SUMER. There are several possible explanations for this
non-consistency. First, the rest wavelength determined by us might not be
accurate. The TR is known to be much more dynamic than the corona, and thus
the TR emission is more likely influenced by such plasma dynamics as related
with spicules. As a result, the average TR line in the off-limb region might
deviate from its rest wavelength. Second, most of the Ne~{\sc{viii}}-related
coronal hole studies were based on observations made in the last solar
minimum, during which coronal holes might have had properties different from
those in the current peculiar solar minimum.

The pattern of the contours enclosing the lowest 20\% percent of the
Doppler values presented in Figure~\ref{fig.1} reveals a clear
dependence on the temperature. In the coronal hole, these contours
are isolated when seen in the upper-TR line Si~{\sc{vii}}. The
contours seem to expand with increasing temperature and finally
merge in the corona at around $\lg(T/\rm{K})=$6.2. This behaviour
agrees with the scenario of solar wind outflow being guided by the
expanding magnetic funnels \citep{Tu2005}, if we assume that lines
with a higher formation temperature are formed in higher atmospheric
layers. Several small funnels which originate from different parts
of the chromospheric network expand with height and finally merge
into a single open field region \citep{Tian2008a}. The nascent solar
wind is guided through these funnel structures.

The temperature variation of the Doppler shift for the averaged profiles in
the quiet Sun (region 2) shows a trend similar to the SUMER results at
$\lg(T/\rm{K})\leq$6.0 \citep{PeterJudge1999,Xia2004}, from red shift in the
middle-TR to blue shift in the upper-TR. Figure~\ref{fig.3} reveals that
above $\lg(T/\rm{K})=$6.0 the averaged profiles in the quiet Sun have very
small blue shifts ($\sim$1.5~km/s), which seem to be independent of the
temperature. This result confirms that our determination of the rest
wavelengths is reasonable, since the quiet-Sun emission at coronal
temperatures usually is assumed to be at rest on average
\citep[e.g.,][]{Milligan2009}.

In the coronal hole, the O~{\sc{v}} line also has a net red shift.
However, from Figure~\ref{fig.3} we can see that the value of the
red shift is smaller in the coronal hole than in the quiet Sun. The
transition from the red shift to the blue shift seems to occur at
around $\lg(T/\rm{K})=$5.6, above which a clear increasing trend of
the blue shift with increasing temperature can be seen in each
selected region of the coronal hole. This blue shift reaches 25~km/s
at around $\lg(T/\rm{K})=$6.3, after having taken the line-of-sight
effect into account. Since emission lines with higher formation
temperatures are usually formed in higher layers, the
temperature-dependent blue shift revealed in Figure~\ref{fig.3} is
likely to be a signature of the initial acceleration of the fast
solar wind.

Based on observations of the Ultraviolet Coronagraph Spectrometer
\citep[UVCS,][]{Kohl1995}, the speed of the nascent fast solar wind
along field lines in the extended corona has been measured by using
the Doppler dimming technique
\citep[e.g.,][]{Giordano2000,Teriaca2003,Gabriel2003,Kohl2006,Antonucci2006,Telloni2007}.
However, in the inner corona, the radial evolution of the outflow
speed has not been well determined. SUMER measurements at off-limb
positions can only provide the line-of-sight component of the
velocity, which is usually very small and largely different from the
real radial outflow speed. SUMER can measure a large fraction of the
real radial outflow speed at disk positions, but the information on
the velocity in the corona is missing due to the lack of coronal
lines in the SUMER spectral range. Our results based on EIS
observations reveal important information on this missing part, and
thus provides further observational constraints on future solar wind
models.

When looking at Figure~\ref{fig.1}, the significant redshifts of the
Fe~{\sc{xii}} and Fe~{\sc{xiii}} lines at the limb appear puzzling
to us, if we assume that non-radial velocities are cancelling each
other at and above the limb. A possible explanation for the redshift
at the limb could be the following. Our observation was made in
October, when the north pole of the solar rotation axis was tilting
towards the Earth. If we simply assume that plasma is flowing
outward along a super-radial expanding funnel structure at the north
limb, then we will see redshift at the side away from the Earth, and
blueshift at the side toward the Earth. Our derived Doppler shifts
will be changed if we assume that the limb emission is at rest.
However, the trend of larger blue shift with increasing temperature
should remain, since the redshift at the limb is larger at higher
temperatures.

\cite{DelZanna2008} studied the temperature-dependent upflows at the boundary
of an active region, which might be related to the origin of the slow solar
wind \citep{Harra2008}. The blue shift at the boundary of the active region
also increases with increasing temperature, and the values of the coronal
Doppler shift in \cite{DelZanna2008} are close to those obtained in the
coronal hole here. In a recent solar wind model presented by
\cite{Cranmer2007}, the outflow speeds for the fast and slow solar wind are
about 30~km/s at the height of 0.2 solar radii, and are not so different
below 0.2 solar radii above the photosphere. Because of the exponentially
decreasing electron density above the photosphere, it is likely that most of
the coronal emission recorded by EIS comes from a certain height range below
0.2 solar radii. Thus, the model results in \cite{Cranmer2007} seem to be
consistent with EIS observations.

Flare and CME related outflows were also found by \cite{Imada2007}
and \cite{Jin2009}. However, these transient outflows seem to start
in the lower TR (formation height of He~{\sc{ii}}), and reach tens
or even more than hundreds of km/s in the corona. Moreover, the
trend of the upflow dependence on temperature dramatically changes
at $\sim$1~MK during flares and CMEs. While both of our
Figure~\ref{fig.3} and the Figure~3 in \cite{DelZanna2008} seem to
suggest a steady acceleration of the solar wind. Thus, behaviors of
the solar wind at its origin and of the CME/flare during initiation
are largely different.

The temperature variation of the difference of Doppler shift between
the coronal hole and quiet Sun as shown in Figure~\ref{fig.3}D is
also consistent with the SUMER result at $\lg(T/\rm{K})\leq$6.0
\citep{Xia2004}. The significant blue shift in the coronal hole with
respect to the quiet Sun suggests that the solar wind outflow starts
in the TR, and corresponding signatures are probably present at
temperatures as low as $\lg(T/\rm{K})=$5.4. Our new result based on
EIS observational data complements the earlier SUMER result, as we
find that the difference of the Doppler shifts between the coronal
hole and the quiet Sun increases with temperature from
$\lg(T/\rm{K})=$5.4 to $\lg(T/\rm{K})=$6.3.

\begin{acknowledgements}
EIS is an instrument onboard {\it Hinode}, a Japanese mission developed and
launched by ISAS/JAXA, with NAOJ as domestic partner and NASA and STFC (UK)
as international partners. It is operated by these agencies in cooperation
with ESA and NSC (Norway). Hui Tian and Chuanyi Tu are supported by the
National Natural Science Foundation of China under contract 40874090.
\end{acknowledgements}

\end{document}